\documentclass{Interspeech2024}

\interspeechcameraready 

\usepackage{multirow}
\usepackage{adjustbox}
\usepackage{subcaption}

\title{Enhancing Speech Emotion Recognition through Segmental Average Pooling of Self-Supervised Learning Features}

\name[affiliation={1}]{Jonghwan}{Hyeon}
\name[affiliation={1}]{Yung-Hwan}{Oh}
\name[affiliation={1}]{Ho-Jin}{Choi}

\address{
  $^1$School of Computing, KAIST, Daejeon, South Korea
}
\email{jonghwanhyeon@kaist.ac.kr, yhoh@kaist.ac.kr, hojinc@kaist.ac.kr}


\begin{document}

\maketitle

\begin{abstract}
Speech Emotion Recognition (SER) analyzes human emotions expressed through speech. Self-supervised learning (SSL) offers a promising approach to SER by learning meaningful representations from a large amount of unlabeled audio data. However, existing SSL-based methods rely on Global Average Pooling (GAP) to represent audio signals, treating speech and non-speech segments equally. This can lead to dilution of informative speech features by irrelevant non-speech information. To address this, the paper proposes Segmental Average Pooling (SAP), which selectively focuses on informative speech segments while ignoring non-speech segments. By applying both GAP and SAP to SSL features, our approach utilizes overall speech signal information from GAP and specific information from SAP, leading to improved SER performance. Experiments show state-of-the-art results on the IEMOCAP for English and superior performance on KEMDy19 for Korean datasets in both unweighted and weighted accuracies.
\end{abstract}

\section{Introduction}

Speech emotion recognition (SER) is an active area of research in the field of speech processing, aiming to automatically recognize the emotional state of a speaker from their speech signal. SER has gained significant attention due to its potential applications in various domains such as human-computer interaction, virtual assistants, and affective computing where understanding the emotional context can greatly enhance the interaction between humans and machines. However, accurately recognizing emotions from speech signals remains a challenging task due to the complex nature of human emotions and the variability of speech signals across different speakers and contexts. 

One of the key challenges in SER is to extract and utilize meaningful and effective features from speech signals for accurate emotion recognition. Traditionally, SER systems rely on handcrafted features, such as Mel-frequency cepstral coefficients (MFCCs), spectral features, and prosody features, which are designed to capture specific aspects of speech signals. However, these features are limited in their ability to capture the complex and dynamic nature of emotions conveyed through speech because they do not capture the higher-level abstractions that are essential for emotion recognition.

Recently, self-supervised learning (SSL) has gained significant success in the natural language processing field, where models are trained on large amounts of unlabeled text data and learn to capture complex contextual relationships between words and phrases. Inspired by this success, researchers have explored the use of SSL models to extract more abstract and informative features from speech signals. These models are trained on large amounts of unlabeled speech data and can learn to capture a wide range of speech characteristics, including phonetic, syntactic, and semantic information potentially capturing more comprehensive and contextualized information. 

Meanwhile, speech signals inherently vary in length, resulting in features extracted from SSL models also having variable lengths. To leverage these variable-length SSL features in machine learning models, which typically require fixed-length representations for input, it is essential to transform them into a fixed-length format. The traditional approach for this transformation is to apply Global Average Pooling (GAP) on SSL features across the temporal dimension. However, speech signals primarily consist of two types of segments: speech segments, which convey meaning through words and phrases, and non-speech segments, which consist of silence and background noise. Since GAP treats all segments equally, whether they are speech or non-speech, it can lead to the dilution of informative features extracted from speech segments by irrelevant information contained within non-speech segments. Consequently, this can negatively impact the performance of SER models that use SSL features.

To solve this problem, we propose Segmental Average Pooling (SAP), which focuses only on speech segments of speech signals, while ignoring non-speech segments. By applying both GAP and SAP on SSL features, our proposed model can utilize overall information of the speech signal from the GAP representation and specific information of the speech signal from the SAP representation.

We evaluate our proposed approach on two datasets, IEMOCAP \cite{busso2008iemocap} for English and KEMDy19 \cite{noh2021multi, noh2023emotion} for Korean, using both unweighted and weighted accuracy. We perform the leave-one-speaker-out cross-validation to measure performance independently of speaker characteristics. Our proposed approach, which combines GAP and SAP, achieves better performance on both datasets compared to relying solely on GAP. Furthermore, we achieve state-of-the-art performance on both datasets, demonstrating the effectiveness of our proposed approach.

Our main contributions are as follows:
(1) We propose a novel pooling method, SAP, which focuses only on speech segments and ignores non-speech segments to prevent the dilution of informative features.
(2) We demonstrate that combining GAP and SAP improves the performance of SER models that use SSL features.
(3) We achieve state-of-the-art performance on the IEMOCAP for English and superior performance on the KEMDy19 for Korean using our proposed approach.
\section{Related Works}
\subsection{Speech emotion recognition}
Speech emotion recognition (SER) is an active area of research that aims to detect the emotional state of a speaker based on characteristics of their speech signal. Over the years, various machine learning techniques have been employed on different types of acoustic features extracted from the speech. Early SER systems utilized Gaussian Mixture Models (GMMs) trained on low-level descriptors such as pitch, energy, and Mel-Frequency Cepstral Coefficients (MFCCs) \cite{lee2005toward, el2007speech, tang2009emotion}.

With the advent of deep learning, neural network architectures such as convolutional neural networks (CNNs) \cite{tzirakis2018end} and long short-term memory (LSTM) networks \cite{huang2016attention} have achieved state-of-the-art performance by learning discriminative feature representations directly from the raw audio. Some studies have also explored using auxiliary modalities like text transcripts \cite{yoon2018multimodal} or visual facial expressions \cite{go2003emotion, busso2004analysis} to complement the acoustic speech data. Recently, advances in self-supervised learning \cite{devlin2018bert, liu2019roberta} and transformer architectures \cite{wolf2020transformers} have further enhanced SER performance. However, challenges still persist in real-world deployment scenarios with noisy inputs and underrepresented emotion classes.

\subsection{Self-supervised learning}
Self-supervised learning (SSL) has emerged as a powerful technique that leverages large amounts of unlabeled data to train models through carefully designed self-supervision tasks. This pre-training process enables models to learn the intrinsic characteristics and patterns present in the data, acquiring rich, contextualized representations. These learned representations can then be effectively fine-tuned for various downstream tasks using a relatively small amount of labeled data and a limited number of training epochs, achieving competitive or even state-of-the-art performance.

In recent years, various SSL models have been introduced in the speech processing field, such as Wav2Vec 2.0 \cite{baevski2020wav2vec}, HuBERT \cite{hsu2021hubert}, and WavLM \cite{chen2022wavlm}. These models have demonstrated significant improvements in performance compared to previous approaches across a wide range of downstream speech tasks, including automatic speech recognition, keyword spotting, speaker identification, as shown in \cite{yang2021superb}.

\section{Proposed Approach}
\begin{figure}[t]
  \centering
  \includegraphics[width=\linewidth]{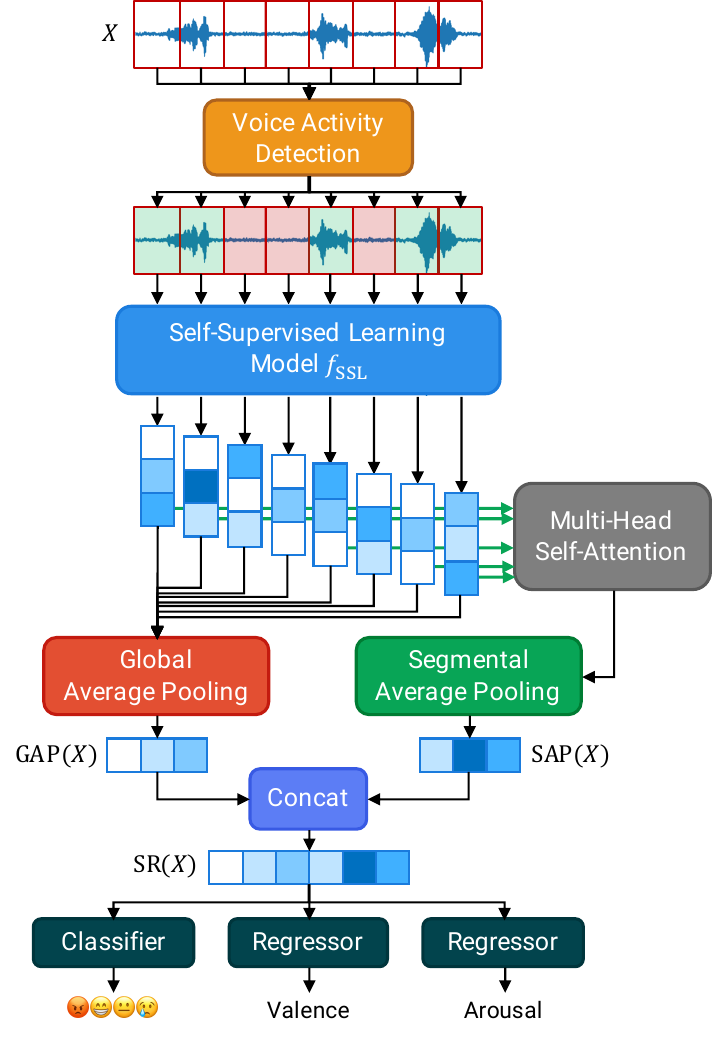}
  \caption{An overall architecture of our proposed approach}
  \label{figure:architecture}
\end{figure}

In this paper, we propose a novel approach to enhance speech emotion recognition (SER) by applying both Global Average Pooling (GAP) and Segmental Average Pooling (SAP) on self-supervised learning (SSL) features. An overall architecture of our proposed approach is illustrated in Figure \ref{figure:architecture}.

\subsection{Self-supervised learning features}
SSL models, which are pre-trained on large-scale audio data, allow us to obtain contextualized speech features directly from a raw speech signal of a given utterance. Let $X$ be a raw speech signal of an utterance $u$. To feed $X$ into SSL models, $X$ is first divided into a sequence of frames $X=(x_1, x_2, \dots, x_T)$, where $x_i \in \mathbb{R}^w$ represents the $i$-th frame of the utterance $u$, and $T$ is the number of frames determined by the length of the raw speech signal, the window size $w$ and the stride $s$. Given a pre-trained SSL model $f_\text{SSL}(\cdot)$,
\begin{equation}
    f_\text{SSL}(X) = [c_1, c_2, \dots, c_T]
\end{equation}
where $c_i \in \mathbb{R}^{d_{\text{SSL}}}$ is a contextualized high-level speech feature for a frame $x_i$, and $d_{\text{SSL}}$ is the dimension of the speech feature from $f_\text{SSL}(\cdot)$.

\subsection{Global Average Pooling}
Since the primary objective of SER is to recognize the emotion conveyed by the entire utterance $u$, rather than the emotion at the individual frame level $x_i$, it becomes crucial to aggregate these frame-level speech features obtained by $f_\text{SSL}(\cdot)$ into a single utterance-level speech feature. 
A traditional approach to achieve this aggregation is to apply Global Average Pooling (GAP) across the temporal dimension of these frame-level speech features as follows:
\begin{equation} \label{equation:gap}
    \text{GAP}(X) = \frac{1}{|f_\text{SSL}(X)|} \sum_{c_i \in f_\text{SSL}(X)}{c_i}
\end{equation}

\subsection{Segmental Average Pooling}
Speech signals primarily consist of two types of segments: speech segments, which convey meaning through words and phrases, and non-speech segments, which consist of silence and background noise. However, in equation \ref{equation:gap}, GAP treats all frames equally, regardless of whether they are speech segments or non-speech segments. This can lead to the dilution of informative features extracted from speech segments by irrelevant information contained within non-speech segments. Consequently, this may negatively impact the performance of SER models that utilize SSL features.

To address this issue, we propose Segmental Average Pooling (SAP), which focuses only on speech segments of speech signals, while ignoring non-speech segments. This selective approach ensures that only informative features extracted from speech segments contribute to the final utterance-level feature. To define $\text{SAP}(\cdot)$, it is necessary to determine whether a given frame contains speech. For this purpose, we utilize the voice activity detection (VAD) algorithm.\footnote{Our proposed approach employs the voice activity detection algorithm developed by Google for the WebRTC project.}
\begin{equation} \label{equation:vad}
    \text{VAD}(x) =
        \begin{cases}
        1 & \text{if } x \text{ contains speech} \\
        0 & \text{otherwise}
        \end{cases}
\end{equation}

Using $\text{VAD}(\cdot)$, we collect frame-level SSL features only from speech segments. Then, we apply multi-head self-attention (MHSA) to these features to capture additional relationships among speech segments as described below:
\begin{equation}
    \begin{gathered}
        g(X) = [c_i | c_i \in f_{\text{SSL}}(X), \text{VAD}(x_i) = 1] \\
        h(X) = \text{MHSA}(g(X))    
    \end{gathered}
\end{equation}
As a result, we define $\text{SAP}(\cdot)$ as follows:
\begin{equation} \label{equation:sap}
    \text{SAP}(X) = \frac{1}{|h(X)|} \sum_{h_i \in h(X)}{h_i}
\end{equation}

\subsection{Combining GAP and SAP}
Our proposed approach leverages the complementary strengths of both GAP and SAP representations. $\text{GAP}(\cdot)$ captures the overall, global information of the speech signal, providing a broad context that includes the average characteristics of the entire signal. Conversely, $\text{SAP}(\cdot)$ focuses on specific, salient features of the speech signal, which are crucial for accurately distinguishing between nuanced phonetic elements or speech characteristics. Therefore, we define the final speech representation, $\text{SR}(\cdot)$, as the concatenation of $\text{GAP}(\cdot)$ and $\text{SAP}(\cdot)$:
\begin{equation}
    \text{SR}(X) = \text{Concat}(\text{GAP}(X), \text{SAP}(X))
\end{equation}

\subsection{Multi-task learning}
Human emotions are complex, and available SER datasets are often limited in size. This necessitates a strategy that maximizes the information extracted from each data sample. To address this challenge, we adopt a multi-task learning (MTL) approach, which aims to concurrently predict both continuous and discrete emotions. The total loss $L$ is defined as:
\begin{equation}
L = \alpha L_\text{discrete} + \beta L_\text{valence} + \gamma L_\text{arousal}
\end{equation}
where $L_\text{discrete}$ is the weighted cross-entropy loss \footnote{Class weights are calculated using the label distribution in the training dataset.} for predicting discrete emotions, and $L_\text{valence}$ and $L_\text{arousal}$ are the mean absolute error losses for predicting continuous valence and arousal emotions, respectively. The coefficients $\alpha$, $\beta$, and $\gamma$ balance the contribution of each loss component to the total loss.

\section{Experiments}
We conduct our experiments on the Interactive Emotional Dyadic Motion Capture (IEMOCAP) \cite{busso2008iemocap} dataset for English and the Korean Emotion Multimodal Database in 2019 (KEMDy19) \cite{noh2021multi, noh2023emotion} for Korean. 

The IEMOCAP dataset consists of five sessions in total, where each session features one male and one female speaker engaged in a conversation. Similar to previous studies, the utterances labeled as "excited" are merged into "happy", and only four emotion classes \{angry, happy, neutral, and sad\} are considered. As a result, the number of utterances representing angry, happy, neutral, and sad are 1103, 1636, 1708, and 1084, respectively. To compare our performance with existing studies under the same conditions, we employ the leave-one-speaker-out 10-fold cross-validation approach, where 8, 1, 1 folds are used as training, validation, and test sets, respectively.

Similarly, the KEMDy19 dataset includes twenty sessions, with each session featuring one male and one female speaker engaged in a conversation. We also consider only four emotion classes \{angry, happy, neutral, sad\}. As a result, the number of utterances representing angry, happy, neutral, and sad are 1530, 1313, 4328, and 773, respectively. To evaluate the performance independently of speaker characteristics, we perform the leave-one-speaker-out 40-fold cross-validation, where 38, 1, 1 folds are used as training, validation and test sets, respectively.

\subsection{Experimental setup}

\noindent\textbf{Evaluation metrics}: 
Following previous studies \cite{yoon2019speech, peng2021efficient, cao2021hierarchical, kim2022improving}, we use unweighted accuracy (UA) and weighted accuracy (WA) as our evaluation criteria. 

\noindent\textbf{Self-supervised learning model}: 
We use WavLM Large \cite{chen2022wavlm} as our self-supervised learning (SSL) model $f_\text{SSL}$ which has achieved competitive performance in the SER task on the SUPERB benchmark \cite{yang2021superb}. 
According to WavLM Large, it uses the window size $w$ of 25ms and the stride $s$ of 20ms.

\noindent\textbf{Multi-task learning}:
We use $\alpha$, $\beta$ and $\gamma$ as 0.5, 0.25 and 0.25, respectively.

\noindent\textbf{Projection dimension}:
The final speech representation $\text{SR}(\cdot)$ is projected into 32 dimensions before feeding it to the classifier for discrete emotion and the regressors for continuous emotions.

\noindent\textbf{Implementation details}:
Our code is implemented using PyTorch \cite{paszke2019pytorch} and HuggingFace Transformers \cite{wolf2020transformers}. 
Due to limited memory capacity, utterances exceeding 19 seconds in the IEMOCAP dataset and 16 seconds in the KEMDy19 dataset are truncated.
We employ an epoch of 30, a batch size of 64, a learning rate of 3e-5, a warm-up ratio of 0.1, and the cosine learning rate scheduler. 
Additionally, we utilize early stopping with a patience of 5, monitoring the total loss $L$ on a validation set.
Our model has approximately 316M trainable parameters. We conduct our experiments using NVIDIA A100 40GB and the estimated training time is about 8 hours in the IEMOCAP dataset and about 40 hours in the KEMDy19 dataset for each experiment. 

\subsection{Results}
\subsubsection{IEMOCAP}
\begin{table}[h]
\centering
\caption{Performance on IEMOCAP}
\label{table:performance-iemocap}
\begin{adjustbox}{width=\columnwidth}
\begin{tabular}{c|cc|cc}
\toprule
\multirow{2}{*}{Method} & \multicolumn{2}{c|}{UA (\%)} & \multicolumn{2}{c}{WA (\%)} \\
                        & Mean     & 95\% CI           & Mean     & 95\% CI          \\
\midrule
$\text{GAP}(\cdot)$     & 73.87    & (71.34, 76.39)    & 73.27    & (70.90, 75.64)   \\
$\text{SAP}(\cdot)$     & 73.48    & (71.13, 75.83)    & 72.45    & (70.34, 74.57)   \\
$\text{SR}(\cdot)$      & \textbf{75.57}    & (72.25, 78.89)    & \textbf{74.77}    & (72.02, 77.52)   \\
\bottomrule
\end{tabular}
\end{adjustbox}
\end{table}

Table \ref{table:performance-iemocap} presents a comparison of the performance of three different methods on the IEMOCAP dataset, employing a leave-one-speaker-out 10-fold cross-validation setting. Compared to $\text{GAP}(\cdot)$ which is a traditional approach for aggregating SSL features, our proposed method, $\text{SR}(\cdot)$, which combines $\text{GAP}(\cdot)$ and $\text{SAP}(\cdot)$, demonstrates superior performance on both UA and WA. However, we observe that $\text{SAP}(\cdot)$ alone does not show an improvement in performance. This indicates that the overall information of the speech signal remains important for accurately recognizing emotions conveyed through speech signals. 

\begin{table}[t]
\centering
\caption{Performance comparison with others on IEMOCAP}
\label{table:performance-comparison-with-others-iemocap}
\begin{tabular}{l|c|c}
\toprule
Model                                       & UA (\%)        & WA (\%)        \\
\midrule
DRN-MHSA \cite{li2019dilated}               & 67.40          & -              \\ 
audio-BRE \cite{yoon2019speech}             & 65.20          & 64.60          \\ 
Audio-CNN-xvector \cite{peng2021efficient}  &68.40           & 66.60          \\ 
HNSD \cite{cao2021hierarchical}             & 72.50          & 70.50          \\ 
MHSA-FACA \cite{kim2022improving}           & 72.83          & 72.01          \\ 
SCL-\textit{k}NN \cite{wang23q_interspeech} & 75.14          & 74.13          \\
\midrule
\textbf{Propposed}                          & \textbf{75.57} & \textbf{74.77} \\ 
\bottomrule
\end{tabular}
\end{table}

Table \ref{table:performance-comparison-with-others-iemocap} presents a comparison of our proposed method with recent state-of-the-art (SOTA) approaches. For a fair comparison, we only consider previous works performing a leave-one-speaker-out 10-fold cross-validation. The results show that our proposed approach, which combines both $\text{GAP}(\cdot)$ and $\text{SAP}(\cdot)$, achieves a relative improvement of 0.43\% and 0.64\% on UA and WA, respectively, compared to other SOTA approaches, demonstrating the effectiveness of our method.

\subsubsection{KEMDy19}
\begin{table}[h]
\centering
\caption{Performance on KEMDy19}
\label{table:performance-kemdy19}
\begin{adjustbox}{width=\columnwidth}
\begin{tabular}{c|cc|cc}
\toprule
\multirow{2}{*}{Method} & \multicolumn{2}{c|}{UA (\%)} & \multicolumn{2}{c}{WA (\%)} \\
                        & Mean     & 95\% CI           & Mean     & 95\% CI          \\
\midrule
$\text{GAP}(\cdot)$     & 64.34    & (62.62, 66.05)    & 66.62    & (64.95, 68.28)   \\
$\text{SAP}(\cdot)$     & 65.72    & (64.32, 67.12)    & 67.75    & (66.59, 68.92)   \\
$\text{SR}(\cdot)$      & 66.26    & (64.59, 67.93)    & 68.27    & (66.82, 69.72)   \\
\bottomrule
\end{tabular}
\end{adjustbox}
\end{table}

Table \ref{table:performance-kemdy19} presents a comparison of the performance of three different methods on the KEMDy19 dataset, employing a leave-one-speaker-out 40-fold cross-validation setting. Similar to the findings with the IEMOCAP dataset, we observe that our proposed approach, $\text{SR}(\cdot)$, shows superior performance in both UA and WA compared to $\text{GAP}(\cdot)$. Since this paper is the first to conduct a leave-one-speaker-out 40-fold cross-validation to evaluate performance independently of speaker characteristics, we are unable to compare our results directly with previous works fairly. However, the performance improvements from the proposed approach, $\text{SR}(\cdot)$, compared to the traditional approach, $\text{GAP}(\cdot)$, demonstrate the effectiveness of our proposed method.

\begin{figure}[h]
    \begin{subfigure}[b]{0.49\columnwidth}
        \includegraphics[width=\linewidth]{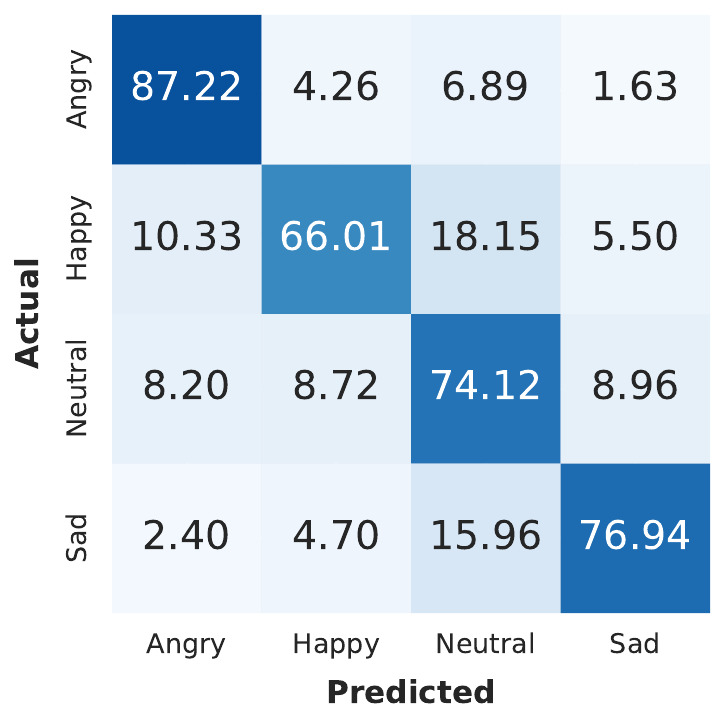}
        \caption{IEMOCAP}
        \label{figure:confusion-matrix-iemocap}
    \end{subfigure}
    \hfill 
    \begin{subfigure}[b]{0.49\columnwidth}
        \includegraphics[width=\linewidth]{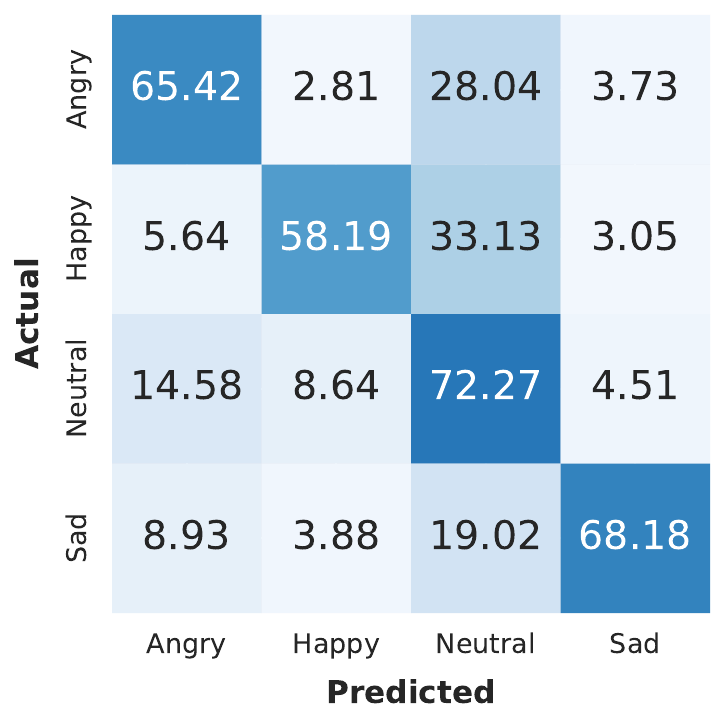}
        \caption{KEMDy19}
        \label{figure:confusion-matrix-kemdy19}
    \end{subfigure}
    \caption{Confusion matrix on IEMOCAP and KEMDy19}
    \label{figure:confusion-matrix}
\end{figure}

Figure \ref{figure:confusion-matrix} shows the confusion matrix generated from our proposed method. According to this matrix, our proposed approach achieves highest accuracy in the angry class on the IEMOCAP dataset and in the neutral class on the KEMDy19 dataset. In contrast, our approach exhibits the lowest accuracy in the happy class on both datasets.
\section{Conclusion}
In this paper, we propose a novel Segmental Average Pooling (SAP) method designed to enhance speech emotion recognition by effectively utilizing self-supervised learning (SSL) speech features. SAP selectively focuses on informative speech segments while ignoring non-speech segments like silence and background noise. By combining SAP with Global Average Pooling (GAP), our approach leverages both overall information from the entire speech signal through the GAP representation and specific information from speech segments through the SAP representation. Our experimental results on two datasets, the IEMOCAP for English and the KEMDy19 for Korean, demonstrate that our proposed approach achieves superior performance compared to relying solely on GAP. Notably, our proposed method achieves state-of-the-art performance on the IEMOCAP dataset and highly competitive results on the KEMDy19 dataset, highlighting its effectiveness in capturing the complex and dynamic nature of emotions conveyed through speech.

\bibliographystyle{IEEEtran}
\bibliography{bibliography}

\end{document}